\documentclass[aps, pra, preprint, groupedaddress, amsfonts,
               amsmath, amssymb, showpacs, nofootinbib]{revtex4-1}
\usepackage{microtype}
\usepackage{graphicx}
\usepackage{epstopdf}
\usepackage[utf8]{inputenc}
\usepackage[T1]{fontenc}
\usepackage[usenames,dvipsnames]{xcolor}
\usepackage{hyperref}

\usepackage{color}

\begin{document}
\title{Net electron capture in collisions of multiply charged projectiles with biologically relevant molecules}

\author{Hans J\"urgen L\"udde}
\email[]{luedde@itp.uni-frankfurt.de}
\affiliation{Center for Scientific Computing, Goethe-Universit\"at, D-60438 Frankfurt, Germany} 

\author{Alba Jorge}
\email[]{albama@yorku.ca}
\affiliation{Department of Physics and Astronomy, York University, Toronto, Ontario M3J 1P3, Canada}

\author{Marko Horbatsch}
\email[]{marko@yorku.ca}
\affiliation{Department of Physics and Astronomy, York University, Toronto, Ontario M3J 1P3, Canada}

\author{Tom Kirchner}  
\email[]{tomk@yorku.ca}
\affiliation{Department of Physics and Astronomy, York University, Toronto, Ontario M3J 1P3, Canada}
\date{\today}
\begin{abstract}
A model for the description of proton collisions from molecules composed of atoms such as hydrogen, carbon, nitrogen, oxygen and phosphorus (H, C, N, O, P)
was recently extended to treat collisions with multiply charged ions with a focus on net ionization. Here we complement the work by focusing on net capture. 
The ion-atom collisions are computed using the two-center basis generator method.
The atomic net capture cross sections are then used to assemble two models for ion-molecule collisions: an independent atom model (IAM) based on the Bragg additivity rule (labeled IAM-AR), and also the so-called pixel-counting method (IAM-PCM) which introduces dependence on the orientation of the molecule during impact. 
The IAM-PCM leads to significantly reduced capture cross sections relative to IAM-AR 
at low energies, since it takes into account the overlap of effective atomic cross sectional areas. 
We compare our results with available experimental and other theoretical data focusing on water vapor ($\rm  H_2O$), methane ($\rm  CH_4$) and  uracil ($\rm C_4 H_4 N_2 O_2$).
For the water molecule target we also provide results from a classical-trajectory Monte Carlo approach that includes dynamical screening effects
on projectile and target. For small molecules dominated by a many-electron atom, such as carbon in methane, or oxygen in water we find a saturation phenomenon
for higher projectile charges ($Q=3$) and low energies, where the net capture cross section for the molecule is dominated by the net cross section for the many-electron atom,
and the net capture cross section is not proportional to the total number of valence electrons.
\end{abstract}
%
%

\maketitle
\section{Introduction}
\label{intro}
Collisions of multiply charged ions with biologically relevant molecules are recognized as being important for future developments in radiation medicine and related fields.
There are numerous experimental efforts which focus largely on differential electron emission, but total ionization, electron capture, and excitation are also relevant
since they contribute to energy deposition, and stopping power.  A recent example is the study of net ionization in $\rm C^{6+} - CH_4$ collisions~\cite{PhysRevA.101.062708}.

On the theoretical side one finds several methods that attempt to explain the experimental data. Often provided alongside with the experimental work is the 
continuum distorted wave with eikonal initial state (CDW-EIS) method. Differential electron emission can be obtained also directly from a classical-trajectory Monte Carlo (CTMC) method,
which for water molecule targets was enhanced recently to include dynamical screening effects~\cite{PhysRevA.99.062701}.
The CTMC approach takes the (frozen) molecular orientation into account during the collision, and total ionization cross sections, as well as some charge-state
correlated cross sections have been compared to experimental data~\cite{jorge2020multicharged}.

Another approach is to use collision information obtained for the atomic constituents and to combine them into molecular cross sections, most notably independent
atom models (IAM), which either follow the simple Bragg additivity rule, or more sophisticated versions that take the molecular structure of the target into account, and
allow for the fact that the effective cross section should be reduced due to overlap effects. 
This pixel counting method (IAM-PCM) tested originally for proton impact~\cite{hjl16,hjl18,hjl19,hjl19b} was
used recently to investigate scaling behavior of net ionization as a function of projectile charge~\cite{PhysRevA.101.062709}.

The CDW-EIS work also relies on ion-atom collision calculations, and includes some molecular effects
on the basis of a Mulliken population analysis. 
For the ionization problem we found that CDW-EIS and CTMC calculations with frozen potentials generally are 
close to IAM-AR results, while the CTMC calculations with dynamical screening~\cite{jorge2020multicharged} are closer to IAM-PCM absolute cross sections~\cite{PhysRevA.101.062709}, 
which are lower in the vicinity of the ionization maximum and  merge with IAM-AR results only at very high energies when the projectile charge is high, i.e., $Q>1$.
The reduction in ionization cross sections (whether in the IAM-PCM results versus IAM-AR or CTMC with versus without dynamical screening) can easily reach a factor of two.

In the present paper we focus on an analogous problem at lower collision energies, namely the net capture problem. Here the ion-atom cross sections grow quickly
with projectile charge, and the overlap and orientation problem again becomes important for the molecular targets. The experimental situation is more scarce in the case
of capture as compared to ionization, and therefore theoretical support is even more important.

The paper is organized as follows. In Sect.~\ref{sec:model} we introduce the theoretical basis for the current work. Sect.~\ref{sec:model1} presents new results for
two-center basis generator method
(TC-BGM) ion-atom calculations for $\rm He^{2+}$ and $\rm Li^{3+}$ projectiles; for atomic hydrogen targets the results are compared with theory and experiment.
In Sect.~\ref{sec:model2} we provide a summary of the IAM-PCM methodology and briefly describe the CTMC time-dependent mean field approach
which is applied to the water molecule target.  Sect.~\ref{sec:expt} serves to provide a detailed comparison with experimental data and a selection
of other theoretical work: in Sect.~\ref{sec:expt1} we present results for
water, where we also compare with CTMC results with and without dynamical screening; 
in Sect.~\ref{sec:expt2} we present results for methane for which we demonstrate a saturation effect in the case of projectile charges $Q=3$, 
and  Sect.~\ref{sec:expt3} contains results for uracil.
The paper ends with a few concluding remarks in Sect.~\ref{sec:conclusions}.
Atomic units, characterized by $\hbar=m_e=e=4\pi\epsilon_0=1$, are used unless otherwise stated.

\section{Model}
\label{sec:model}
\subsection{Ion-atom collisions}
\label{sec:model1}

\begin{figure}
\begin{center}$
\begin{array}{ccc}
\resizebox{0.38\textwidth}{!}{\includegraphics{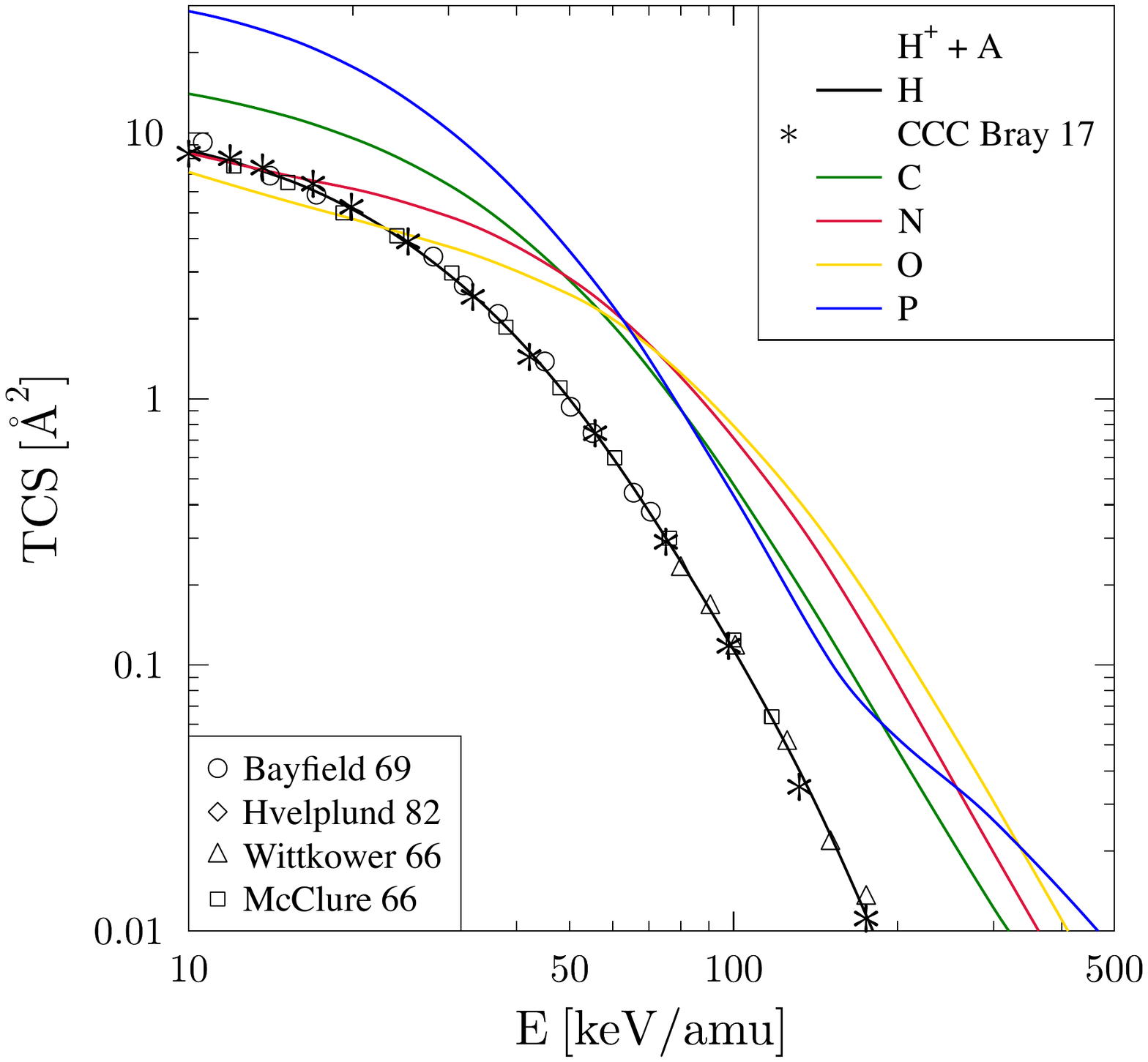}}{\hspace{-1. truecm}}&
\resizebox{0.38\textwidth}{!}{\includegraphics{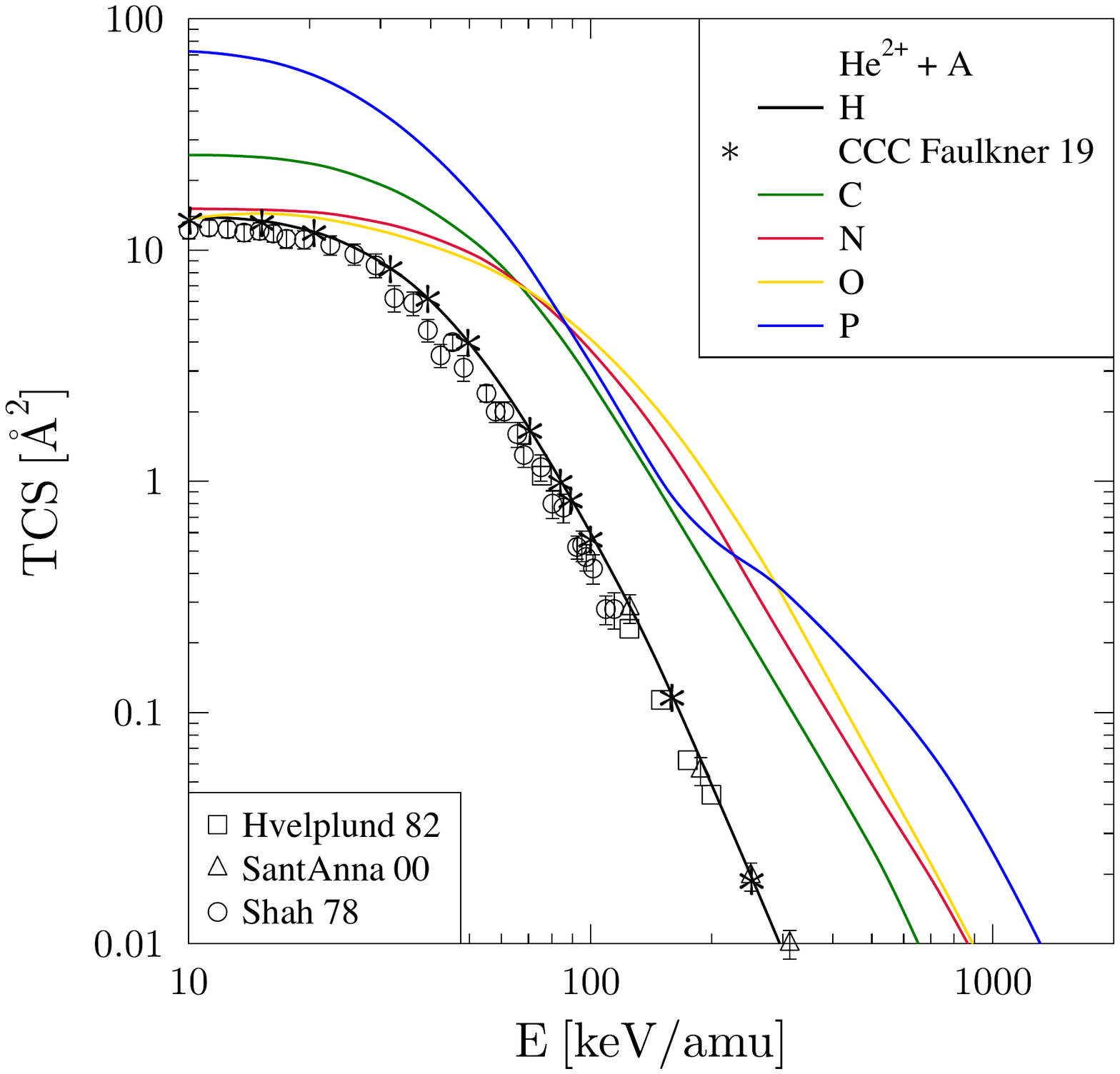}}{\hspace{-1. truecm}}&
\resizebox{0.38\textwidth}{!}{\includegraphics{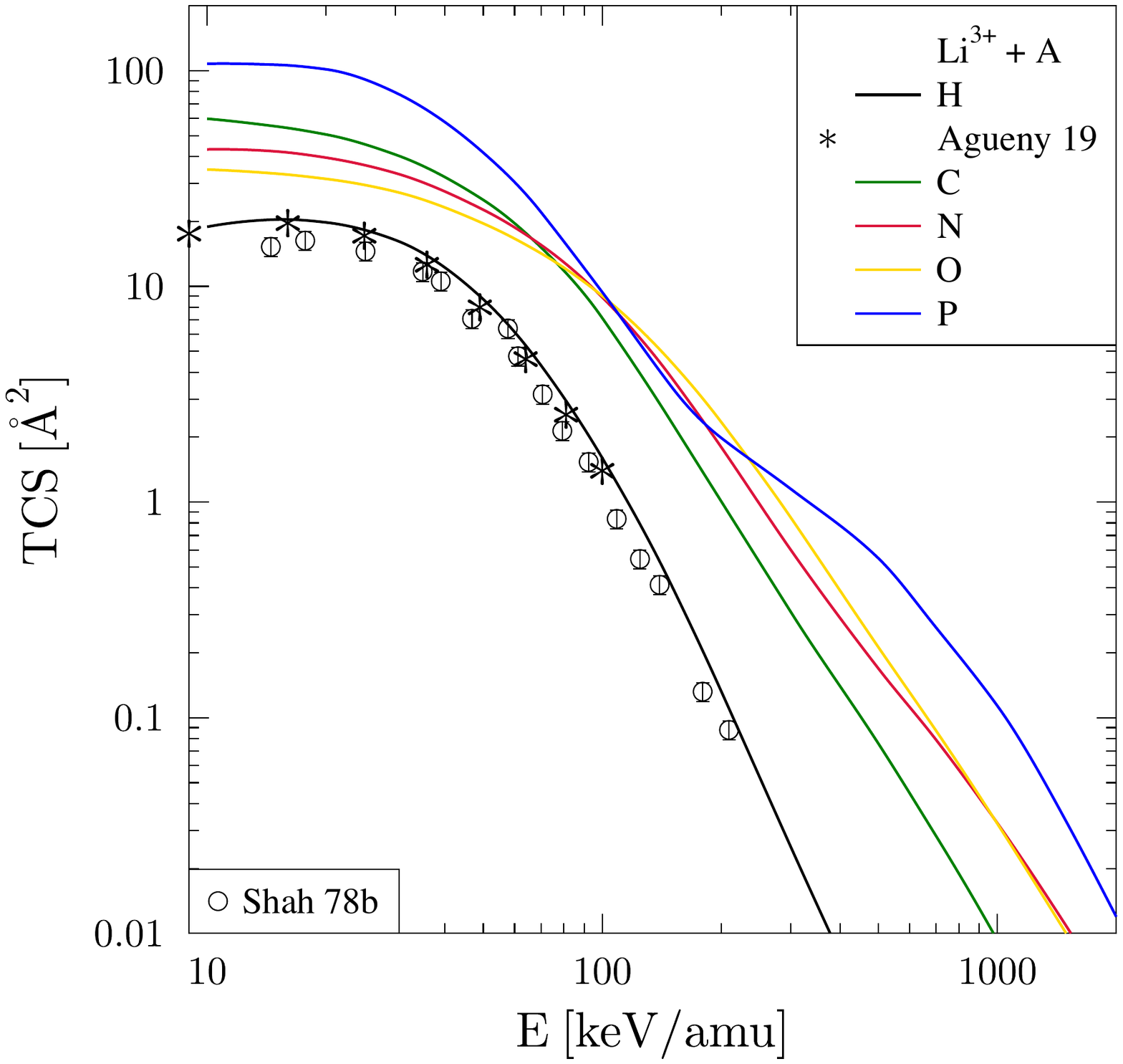}}
\end{array}$
\caption{%
In the left panel the total (net) capture cross sections for protons colliding with atoms H, C, N, O, P is shown (solid lines) as calculated with the TC-BGM
using basis sets described in the text, the middle panel is for He${}^{2+}$ projectiles,
while the right panel is for the case of   Li${}^{3+}$ projectiles. For atomic hydrogen targets and proton and  He${}^{2+}$ projectiles the results are compared
with the convergent close coupling calculation of Ref.~\cite{Bray_2017,Faulkner_2019} respectively. For  Li${}^{3+}$ impact we compare with data from Ref.~\cite{AGUENY2019}.
For proton impact the experimental data are from Ref.~\cite{PhysRev.148.47, PhysRev.185.105,Wittkower_1966,Hvelplund_1982}, for alpha particle impact from Ref.~\cite{Shah_1978,Hvelplund_1982,PhysRevA.61.052717},
and for Li${}^{3+} -  \rm H$ collisions from Ref.~\cite{Shah_1978b}.
}
\label{fig:Abb1}
\end{center}
\end{figure}

The independent atom models rely on state-of-the-art ion-atom collision calculations. In Fig.~\ref{fig:Abb1} we show the results of the two-center basis generator method
(TC-BGM) calculations for atomic targets that form the constituents of biologically relevant molecules (hydrogen, carbon, nitrogen, oxygen, and phosphorus).
Experimental verification is only available for atomic hydrogen, as is theoretical confirmation by other state-of-the-art methods, namely the convergent close coupling (CCC)
approach~\cite{Bray_2017,Faulkner_2019}, and a two-center atomic orbital expansion method based on Gaussian-type orbitals~\cite{AGUENY2019}.

The present TC-BGM ion-atom calculations are as described in Ref.~\cite{PhysRevA.101.062709}: for proton impact a projectile potential $W_P$ hierarchy
is obtained on the basis of explicitly including shells up to principal quantum number $n=4$ on projectile and target, for $Q=2$ projectiles
we used $n_P=6$ and  $n_T=5$, while for $Q=3$ the explicitly included basis was expanded to $n_P=7$ and  $n_T=5$. These explicitly included
states are complemented by states that represent the continuum using the BGM approach~\cite{hjl18,tcbgm}.

For the many-electron atoms the theoretical modelling is at the level of exchange-only density functional theory using the optimized potential method.
Therefore, electron correlation effects are not included in these calculations~\cite{PhysRevA.47.2800,PhysRevA.14.36}.

The results for proton and alpha particle impact on atomic hydrogen display excellent agreement with experiments and with CCC theory over several orders
of magnitude. For projectile charge $Q=3$, i.e., $\rm Li^{3+}$ projectiles, our results are slightly higher than the experimental data, but are in very good agreement
with the recently reported calculations of Ref.~\cite{AGUENY2019}.  
Not shown in Fig.~\ref{fig:Abb1} for proton and  $\rm He^{2+}$ impact are the calculations of Ref.~\cite{AGUENY2019}.
They agree very well the TC-BGM and CCC results for protons and are
slightly lower  (at the $10 \ \%$ level for energies $\rm 25-100 \ keV/amu$) for $\rm He^{2+}$ projectiles.

The structure of these cross sections is simple.
Notable is the increase in net capture as one reduces the collision energy towards $10\ \rm keV/amu$, with cross section values short of 8 \AA${}^2$
for protons, 12 \AA${}^2$ for $Q=2$, and a value of 20 \AA${}^2$ reached already at $30\ \rm keV/amu$ for bare lithium projectiles.

For the atoms containing more than one electron capture from the outer shells becomes large even for intermediate energies with patterns that
are not totally straightforward when one compares $Q=1,2,3$. Capture from phosphorus reaches 100 \AA${}^2$ for $Q=3$ at low energies, and
electron capture from carbon atoms is also very strong. This provides a background for interesting effects when combining these cross sections
to make predictions for molecular targets.

For phosphorus atoms there are notable structures in the capture cross sections at higher energies which are related to shell effects.
For collision energies below $\rm 200 \ keV/amu$ capture from the M-shell dominates, but at higher energies capture from the L-shell becomes
more important. This was investigated, e.g., in Ref.~\cite{tcbgm} for $\rm p-Na$ collisions.

\subsection{Ion-molecule collisions}
\label{sec:model2}

The present work reports on calculations of several theoretical models. On the one hand we compare two independent atom models: {\it (i)} in the naive case,
namely the additivity rule based model (IAM-AR) the atomic capture cross sections described in Section~\ref{sec:model1} are simply added together
and completely ignore molecular structure;
{\it (ii)} in the more sophisticated pixel counting method (IAM-PCM) molecular structure is introduced by a geometric procedure, which will be briefly reviewed below.
For the water molecule target we also report results from two classical-trajectory models.

The IAM-PCM has been described and illustrated in detail in previous papers, most notably Refs.~\cite{hjl16,hjl19b}. Ref.~\cite{hjl19b} highlights 
the role played by the contributions from the time evolution of occupied orbitals from different atomic species to the net cross sections.
The model is compared there to a methodology employed by the community that uses the CDW-EIS approach, which also uses ion-atom calculations,
and incorporates molecular eigenenergies on the basis of quantum molecular structure calculations while employing complete neglect of differential overlap 
(CNDO), i.e., a Mulliken population analysis~\cite{PhysRevA.62.022701,Quinto_2018}.

In essence the IAM-PCM approach is based on an interpretation of the atomic net cross sections as geometrical areas, e.g., areas that correspond
to the net capture (or net ionization) cross section. Rather than summing up all these areas (as assumed by the additivity rule, i.e., the IAM-AR) a pixel
counting method is used to measure the effective cross-sectional area that emerges when one eliminates overlaps between cross sections from
different atoms encountered by the projectile.
Therefore, the effective molecular cross sectional area is defined as a function of
molecular orientation. The latter is chosen randomly, and is sampled to obtain converged cross sections. A critical discussion of the merits of this procedure (potential
emphasis on atoms encountered first) is found in Ref.~\cite{hjl18}, where it is argued that for net cross sections the method should certainly be appropriate.
An illustration of the method is given below in Section~\ref{sec:expt2} where we observe a saturation behavior of the cross section for $Q \ge 3 $ projectiles
colliding with methane ($\rm CH_4$).

For a detailed description of  the other method for which we show results for water molecule vapor targets, namely the CTMC model we also refer to previous literature.
Calculations with frozen target potential on the basis of a three-center potential are reported in Refs.~\cite{PhysRevA.83.052704,Illescas_2020}. This effective potential used 
in the classical statistical ensemble simulation is drawing information from quantum structure calculations. It yields accepted values for
the orbital energies and in addition to CTMC calculations
was also used for numerical grid solutions of the time-dependent Schr\"odinger equation~\cite{ERREA2015}. An extension of this
model which can be considered a semiclassical ($\hbar=0$) approximation to the quantum problem is presented in Ref.~\cite{PhysRevA.99.062701} where the
potential parameters were allowed to vary dynamically as a function of the average (net) ionization state of the water molecule. This resulted in a substantial reduction of the 
net ionization cross section (by up to a factor of two) compared to the static potential model results. A further extension was carried out recently where the projectile
potential was also allowed to be varied as a function of the average (net) charge state of the projectile~\cite{jorge2020multicharged}.

Both the static-potential CTMC and the dynamically screened model are sensitive to the orientation of the molecule during the collision. Therefore, the comparison
with IAM-PCM and IAM-AR calculations is of great interest. Concerning capture data there is one problem that needs to be addressed in the CTMC approach: for high energies capture from inner shells in target atoms (or molecules) becomes problematic in a classical-trajectory model, because it becomes possible to capture into orbits well below the allowed
1s-level, i.e., orbits with binding energies larger than $Q^2/2$ begin to occur. These classically allowed capture contributions need to be removed from the analysis. For proton
impact ($Q=1$) this correction becomes significant at impact energies of $200 \ \rm keV/amu$ and higher, for $Q>1$ this point moves to higher energies.
The correction procedure is based on a prescription to associate principal quantum numbers $n$ with energy ranges~\cite{Becker_1984}.

\section{Comparison with experiments}
\label{sec:expt}

\subsection{Collisions with water vapor ($\rm  H_2O$)}
\label{sec:expt1}

\begin{figure}
\begin{center}
\resizebox{0.6\textwidth}{!}{\includegraphics{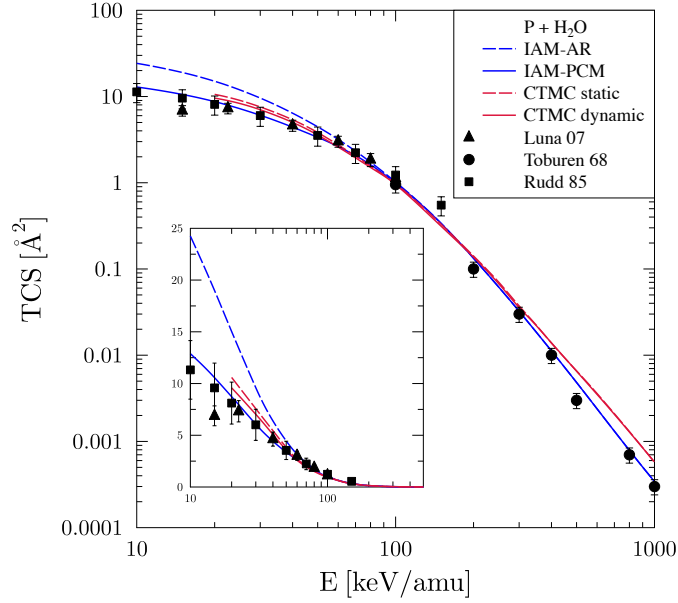}}  
   
\caption{Net electron capture cross section for proton-water collisions. The highest curve (dashed blue) shows the Bragg additivity rule result (IAM-AR), the lowest (solid blue)
curve the IAM-PCM result. In between are the CTMC results, namely the (dashed red) static-potential CTMC, and below it (solid red) the dynamical-screening CTMC result.
The experimental data are from Refs.~\cite{PhysRev.171.114,Rudd85c,Luna07}
}
 \label{fig:Abb2}
\end{center}
\end{figure}

In Fig.~\ref{fig:Abb2} we compare the theoretical model results for proton impact on water vapor with experiment. We note that the IAM-PCM results are corrected
compared to those shown in Refs.~\cite{hjl16,hjl18} (the wrong bond length was applied in those calculations), and they agree now well within error bars
at low energies with the experimental data of Rudd {\it et al.}~\cite{Rudd85c}. The IAM-AR results (which are independent of the molecular structure and remain unchanged) overestimate the low-energy experimental data by up to a factor of two.

The two CTMC model calculations are expected to yield very similar results, since dynamical screening on the projectile is not turned on for $Q=1$, and dynamical
screening on the target is expected to be small. At medium to low energies both CTMC model
calculations fall in between the IAM-PCM and IAM-AR results and are consistent with experiment (at the upper end of their errors). 
The inset which uses a semilogarithmic presentation shows that the model
with dynamical screening is close to IAM-PCM, and still consistent with the experimental data of Ref.~\cite{Rudd85c}.

\begin{figure}
\begin{center}
\resizebox{0.6\textwidth}{!}{\includegraphics{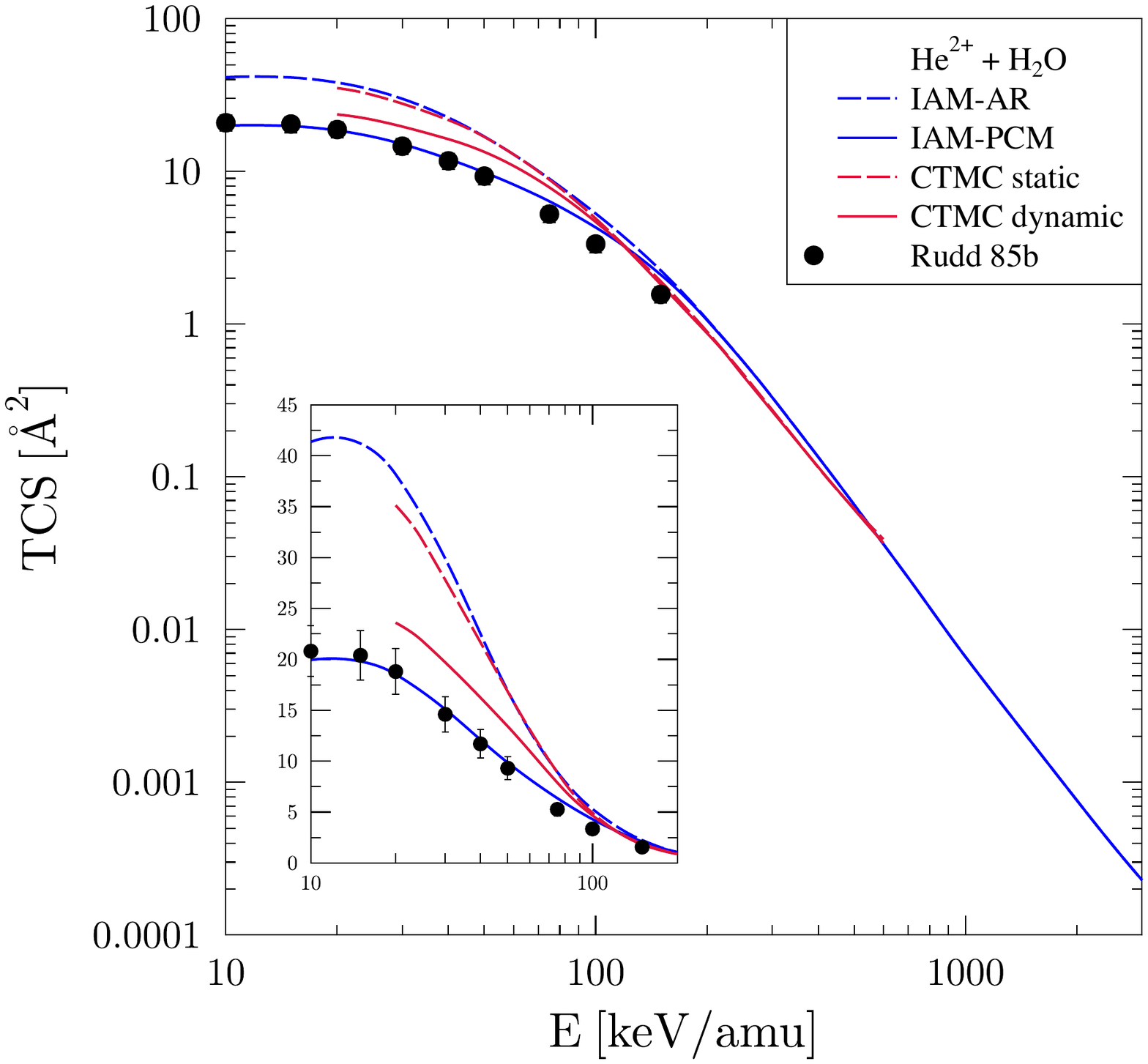}}  
   
\caption{Net electron capture cross section for $\rm He^{2+}$-water collisions. The curves follow the same pattern as described for Fig.~\ref{fig:Abb2}. The experimental data are from Refs.~\cite{PhysRevA.32.2128}.
}
 \label{fig:Abb3}
\end{center}
\end{figure}

In Fig.~\ref{fig:Abb3} we present our results for alpha particle impact on water vapor. The IAM-PCM results follow the trend of the experimental data of Ref.~\cite{PhysRevA.32.2128}
very well, particularly at the lowest energies shown. The IAM-AR model overestimates them by a factor of two at the lowest energies. The CTMC calculations with static
potential are close to the IAM-AR results at intermediate and high energies. The CTMC time-dependent mean-field model calculation, on the other hand, is closer to the IAM-PCM results
and overestimates them by about $20-30 \ \%$ at low to medium energies. For higher energies all models are practically in agreement. The experimental data are obtained by
summing single and double-capture contributions ($\sigma_{\rm net} = \sigma_1 + 2 \sigma_2$), which have error estimates of $12 \ \%$ and $16 \ \%$ respectively.

\begin{figure}
\begin{center}
\resizebox{0.6\textwidth}{!}{\includegraphics{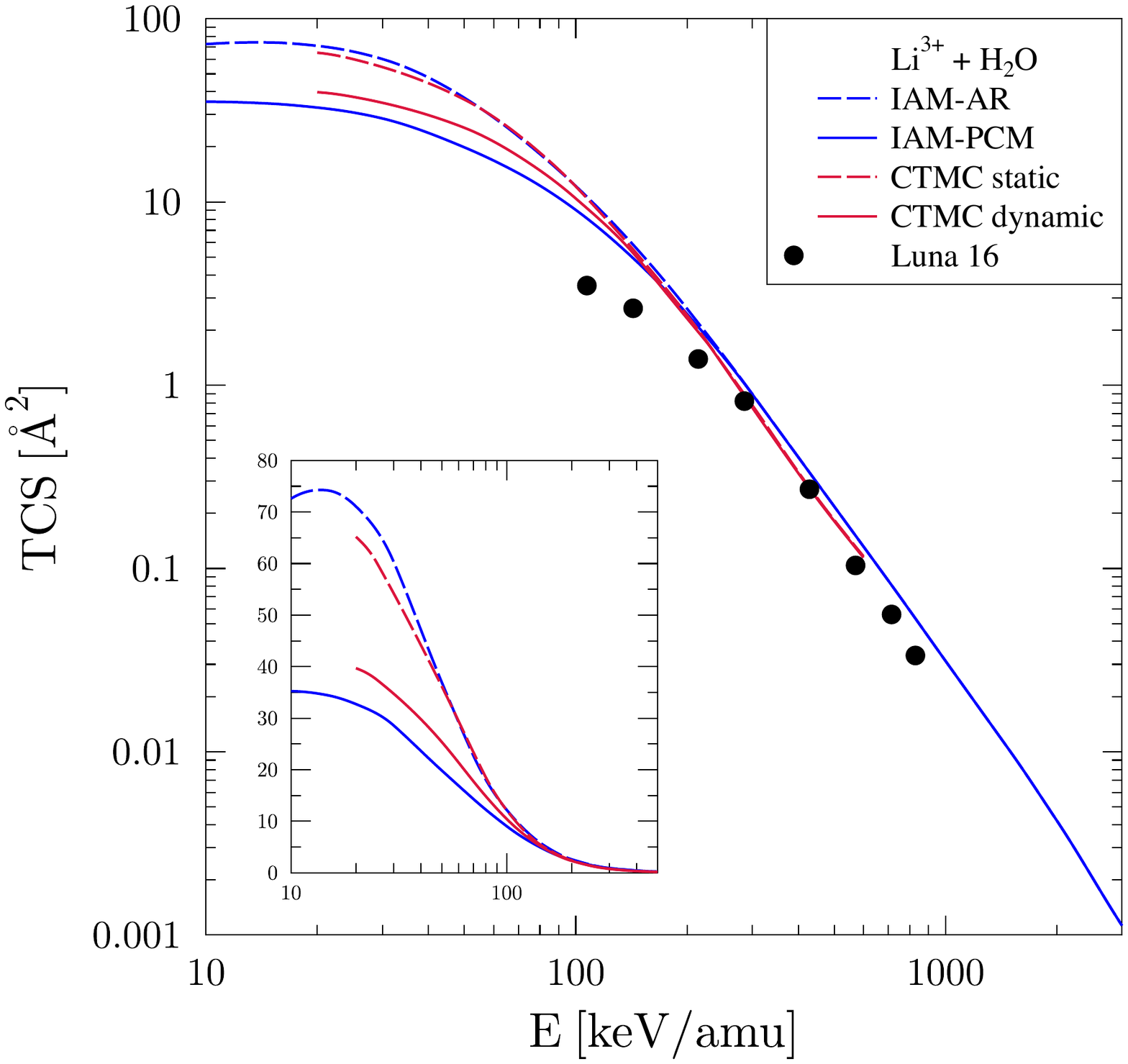}}  
   
\caption{Net electron capture cross section for $\rm Li^{3+}$-water collisions.  The curves follow the same pattern as described for Fig.~\ref{fig:Abb2}. The experimental data are from Ref.~\cite{Luna16}.
}
 \label{fig:Abb4}
\end{center}
\end{figure}

For $\rm Li^{3+}$-water vapor collisions the trend observed for alpha-particle impact continues: the calculated net capture cross sections continue to grow with
projectile charge $Q$, and the gap between IAM-PCM and IAM-AR remains at a factor-of-two increase for the naive Bragg-additivity-rule-based model.
The static-potential CTMC calculations side with this IAM-AR result, while the CTMC time-dependent mean-field calculation is again only $20-30 \ \%$ 
above the IAM-PCM result at low to medium energies.

At energies above $200 \ \rm keV/amu$ the models merge and agree well with the experimental data of Ref.~\cite{Luna16}. The two data points below this energy are
below all calculated values, reaching a factor-of-two discrepancy at $100 \ \rm keV/amu$. This shortfall of the experimental data goes hand-in-hand with
an observed shortfall in double-ionization contributions in this energy range as compared to TC-BGM calculations for pure ionization (cf. Fig.~7 of Ref.~\cite{Luna16})
and transfer ionization (cf. Fig.~8 of Ref.~\cite{Luna16}). Note that these transfer ionization channels contribute both to net capture and net ionization.
This shortfall in ionized electron flux in the experimental data  is difficult to understand in the context of modelling the projectile charge state dependence
both in the IAM-PCM~\cite{PhysRevA.101.062709} and the time-dependent mean-field CTMC model~\cite{jorge2020multicharged}, and thus we can only ask for additional experimental work in this context.

\subsection{Collisions with methane ($\rm  CH_4$)}
\label{sec:expt2}

IAM-PCM and IAM-AR results for net capture of electrons from methane by proton impact were described previously in Ref.~\cite{hjl19b}, where they were
compared with other theoretical works, namely CDW-EIS and CNDO calculations. The CDW-EIS method was applied to higher projectile charges in
Ref.~\cite{Quinto_2018}.

\begin{figure}
\begin{center}
\resizebox{0.6\textwidth}{!}{\includegraphics{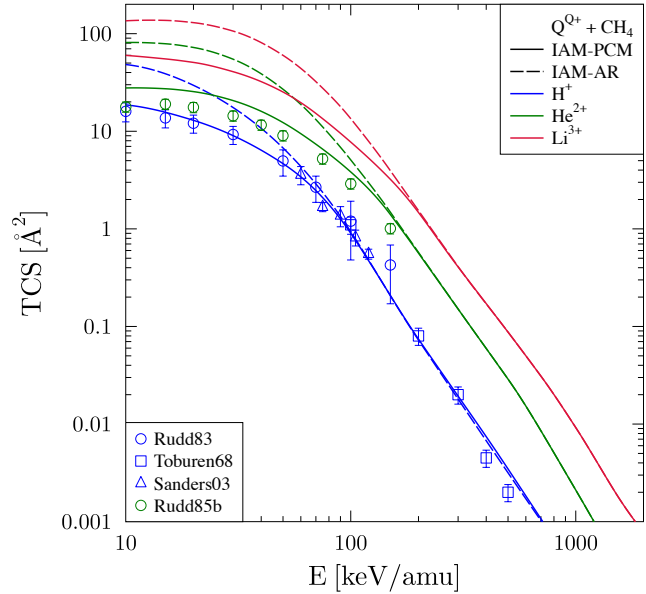}}  
   
\caption{Net electron capture cross section for collisions of protons, alpha particles, and $\rm Li^{3+}$ ions with methane ($\rm CH_4$). 
The dashed curves show the IAM-AR results, while the solid curves are obtained with the IAM-PCM. The curves merge at lower energies for proton impact (shown in blue),
followed by alpha-particle impact (shown in green), and even higher energies for $\rm Li^{3+}$ projectiles (shown in red).
The experimental data are from Refs.~\cite{PhysRev.171.114,PhysRevA.28.3244,PhysRevA.32.2128,Sanders_2003}.
}
 \label{fig:Abb5}
\end{center}
\end{figure}

In Fig.~\ref{fig:Abb5} we compare the net capture cross sections from IAM-AR and IAM-PCM  for projectile charges $Q=1,2,3$.
For the methane target the difference between the two models at the lowest energy reported exceeds a factor of two.
For alpha particle impact the experimental data fall below the present IAM-PCM results, and merge with the proton impact data at $\rm 10 \ keV/amu$
collision energy.

For $\rm Li^{3+}$ projectiles we find that the net capture cross section at low energies becomes 60 \AA${}^2$, which is the same value that is
reached by $\rm Li^{3+}-C$ collisions (cf. Fig.~\ref{fig:Abb1}).
This can be called a saturation behavior in the sense that it is not the total
number of valence electrons that determines the size of net cross sections for high projectile charges. 
We first investigate this phenomenon in detail for methane, and then make some comments further below as to which other target molecules will be affected
in a similar way. Looking at the equivalent comparison of $\rm Li^{3+}-H_2 O$ (Fig.~\ref{fig:Abb4}) vs $\rm Li^{3+}-O$ (Fig.~\ref{fig:Abb1}) collisions  we observe the same 
effect with a common value of about 35 \AA${}^2$.

The saturation behavior in the net capture cross section as one goes to low impact energies and higher projectile charges can be illustrated as
an overlap effect in the IAM-PCM. 
A geometric condition for the saturation behavior in $\rm CH_4$ is obtained from the following consideration:
if $r_C$ and $r_H$ are the radii of the carbon and hydrogen cross-sectional disks with $r_X = \sqrt{\sigma_X/\pi}$  and $b$ the bond length of C-H, then  saturation happens
if $ r_C \ge r_H + b $. For molecules which involve large bond lengths saturation is much less likely to occur.\\

\begin{figure}
\begin{center}$
\begin{array}{ccc}
\resizebox{0.33\textwidth}{!}{\includegraphics{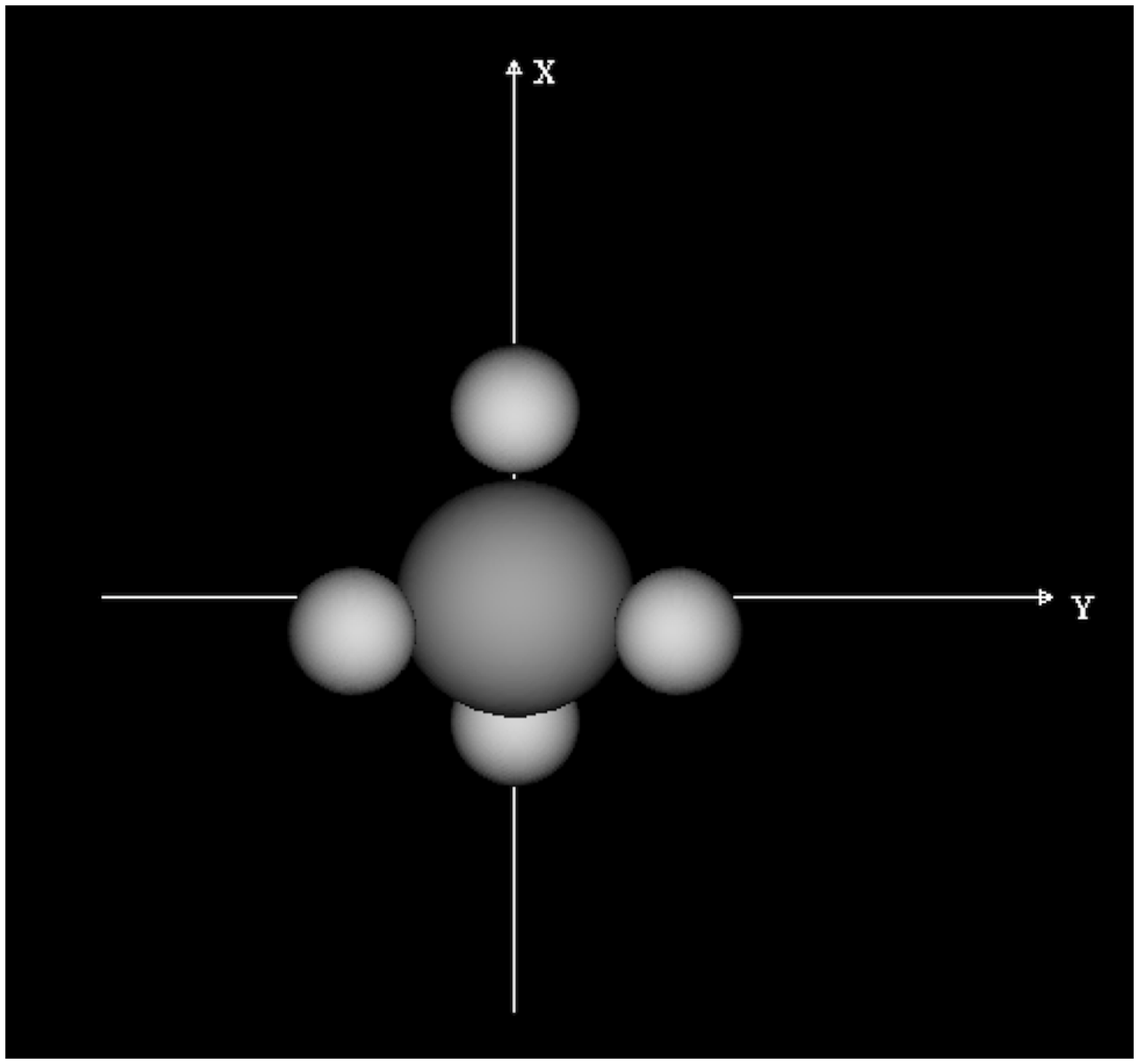}}{\hspace{-1. truecm}}& 
\resizebox{0.33\textwidth}{!}{\includegraphics{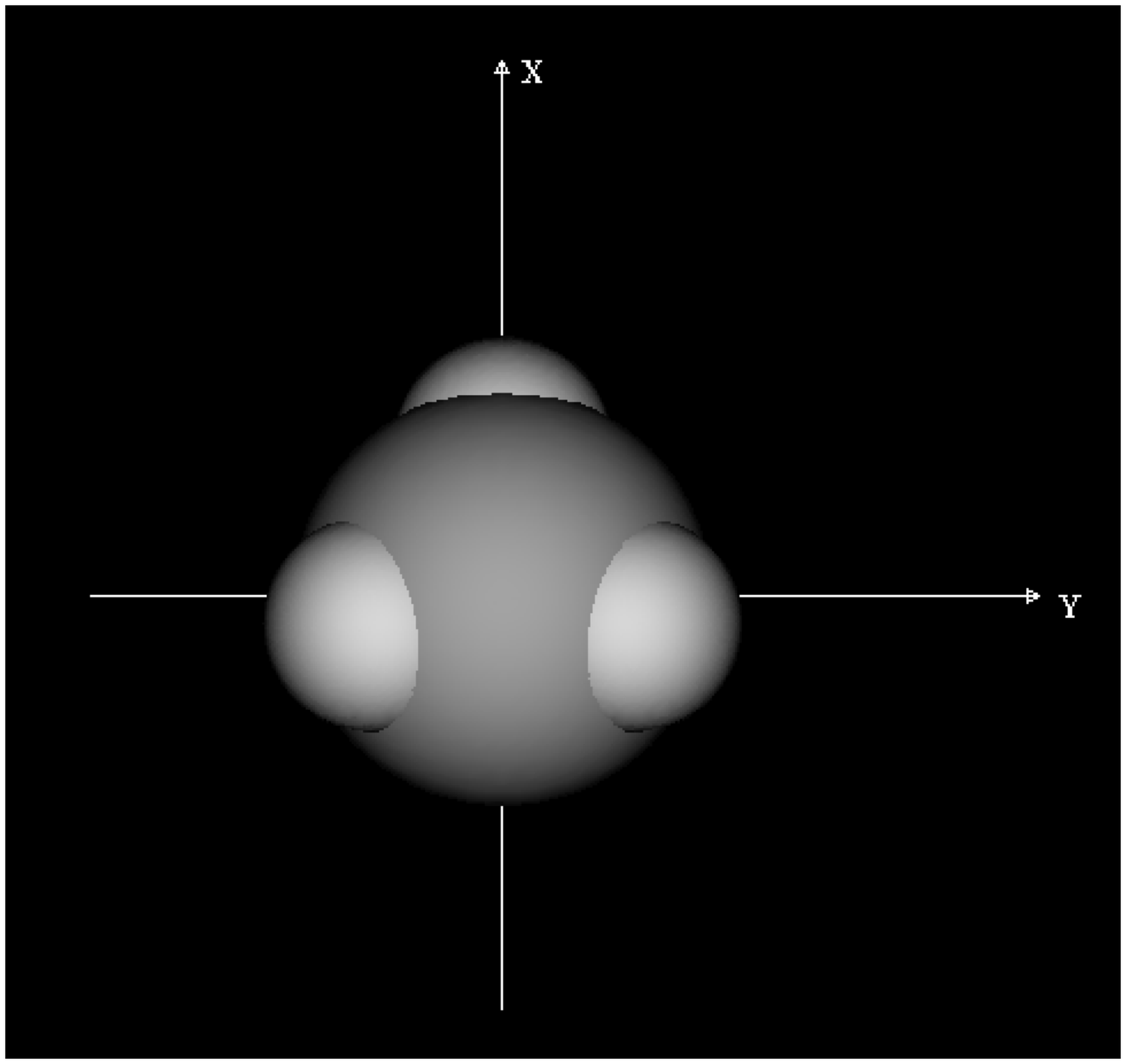}}{\hspace{-1. truecm}}&
\resizebox{0.33\textwidth}{!}{\includegraphics{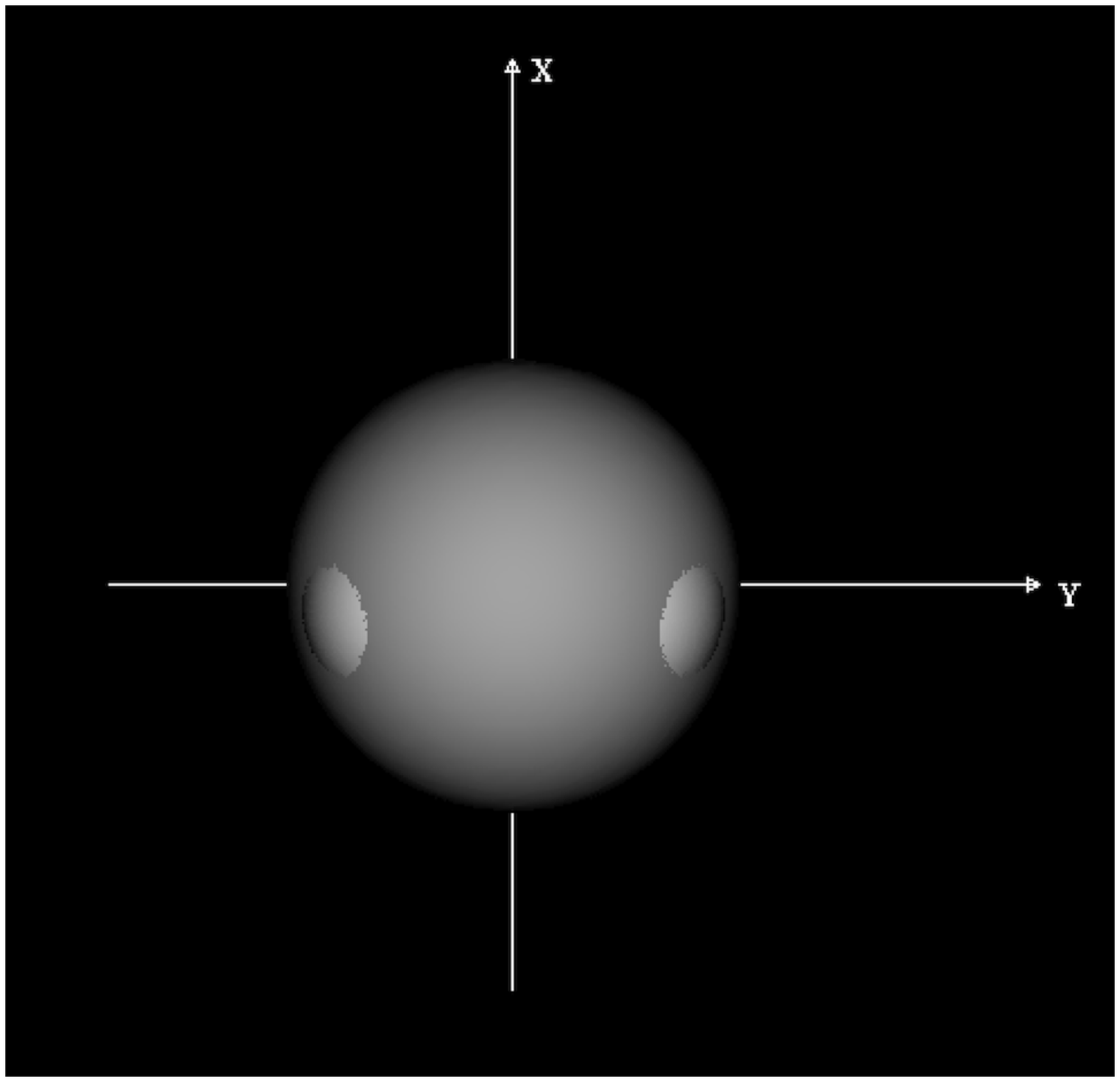}}
\end{array}$
\caption{Demonstration of the overlap effect in IAM-PCM: for a particular orientation the effective net capture cross section is obtained from
overlapping the atomic cross sections for the carbon atom and the four hydrogens at a collision energy of $\rm 70 \ keV/amu$. Left panel for
proton impact ($Q=1$), middle panel for alpha particles ($Q=2$), and right panel for $Q=3$.
}
\label{fig:Abb6}
\end{center}
\end{figure}

As shown in Fig.~\ref{fig:Abb1} net electron capture from carbon atoms for low energies
exceeds net capture from atomic hydrogen by more than a factor of two for $Q=3$ projectiles.
The saturation condition $ r_C \ge r_H + b$ is fulfilled up to $E \approx \ \rm 80 keV/amu$. 
As a result of saturation the IAM-PCM 
cross section becomes independent of the orientation of the methane molecule. This overlap effect is demonstrated for methane in Fig.~\ref{fig:Abb6} and analogous
figures can be drawn for water vapor.

\begin{figure}
\begin{center}
\resizebox{0.6\textwidth}{!}{\includegraphics{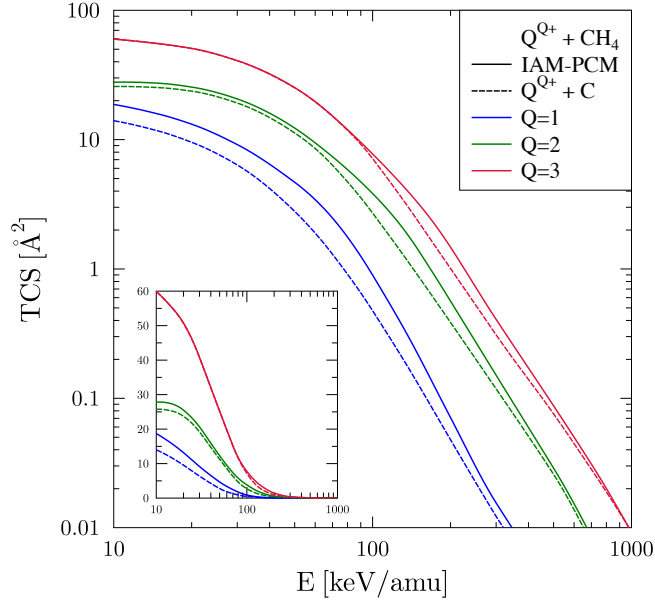}}  
   
\caption{Net electron capture cross sections as functions of impact energy for collisions of protons, alpha particles, and $\rm Li^{3+}$ ions with methane ($\rm CH_4$)  calculated in the IAM-PCM are compared
to net capture from carbon atoms (short dashed lines) as calculated by the TC-BGM (cf. red curves in Fig.~\ref{fig:Abb1}). 
The top curve pair (in red) is for $\rm Li^{3+}$ projectiles, the middle curve pair (in green) for $\rm He^{2+}$ projectiles and the bottom pair (in blue)
for proton impact.
}
 \label{fig:Abb7}
\end{center}
\end{figure}

To further illustrate the saturation phenomenon we show in Fig.~\ref{fig:Abb7} a direct comparison between the IAM-PCM net capture cross sections for ion-methane collisions
with those for the same process involving carbon atoms. With increasing projectile charge $Q$ the two cross sections merge. This feature would be a strong
test for the IAM-PCM if experimental capture cross sections for ion-carbon collisions were available.

Clearly, saturation in the sense that the cross section of just one atom becomes an effective bound for the molecular cross section
is not achievable for large molecules. 
If, however, saturation does occur  (as was observed in methane for carbon atoms vs hydrogen) then one can generalize, namely
one finds $\sigma_{CH_4} = \sigma_{C} = \sigma_{CH_3} = \sigma_{CH_2}$, i.e., the
cross sections of the functional groups $\rm CH_3$ and $\rm CH_2$ are also limited by the cross section of carbon.
For pure hydrocarbons (alkanes, alkenes, alkynes, aromatics, etc.) this means that they can be described as
carbon clusters  with molecular geometry.\\

A similar situation is found for other functional groups of biorelevant molecules. For example, the cross section
for the hydroxyl group (OH) is restricted by the atomic cross section of O, just like the amino group ($\rm NH_2$)  by N
and phosphate ($\rm PO_4$) is limited by the cross section of P. This simplifies the treatment of large biomolecules considerably
in the case of saturation, as they can be approximated by an IAM-PCM model of the dominant atoms (C, N, O, P) at the molecular positions
of  their respective functional groups.\\


   

\subsection{Collisions with uracil ($\rm C_4 H_4 N_2 O_2$)}
\label{sec:expt3}

For large biologically relevant molecules the saturation phenomenon will occur at another level, namely the combinations of large contributor atoms (e.g., 
multiples of carbon, nitrogen)
may again lead to a reduction of the cross sections due to overlap. 
While it may seem that the hydrogen atoms are `covered' up by the overlap effect for high
projectile charges and low collision energies, they do, of course, contribute to the capture process, particularly if there are multiples of them. 
One should keep in mind that net capture includes transfer ionization processes, i.e., for $Q=3$ net capture involves more than three electrons.
The overlap of net
cross sections, i.e., of the geometric areas is naturally dominated by atoms which on account of their valence shells can make big contributions of their own.
The uracil molecule is an interesting candidate to look into the phenomenon.

\begin{figure}
\begin{center}
\resizebox{0.6\textwidth}{!}{\includegraphics{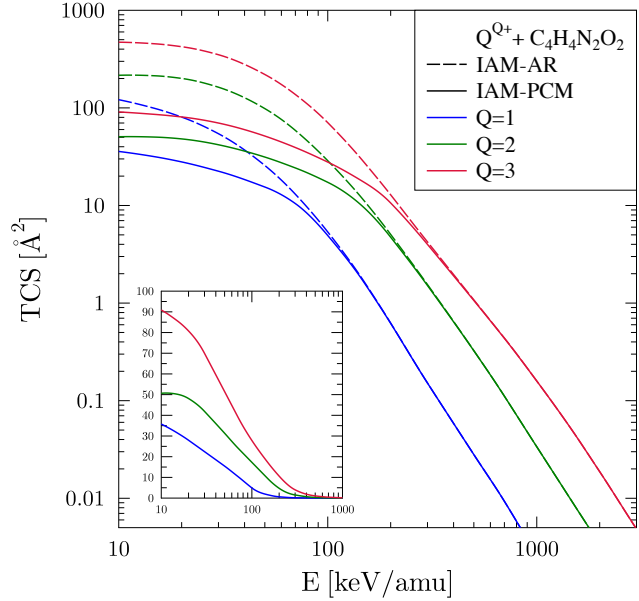}}  
   
\caption{Net electron capture cross section for collisions of protons, alpha particles, and $\rm Li^{3+}$ ions with uracil ($\rm C_4 H_4 N_2 O_2$). 
The dashed curves show the IAM-AR results, while the solid curves are obtained with the IAM-PCM. The curves merge at lower energies for proton impact (shown in blue),
followed by alpha-particle impact (shown in green), and even higher energies for $\rm Li^{3+}$ projectiles (shown in red).
}
 \label{fig:Abb8}
\end{center}
\end{figure}

We begin with a comparison of the net capture cross sections for our three projectiles with charges $Q=1,2,3$ in  Fig.~\ref{fig:Abb8}. The structure of the cross sections
as a function of collision energy is similar to that obtained for methane (cf. Fig.~\ref{fig:Abb5}), although the magnitude of the cross sections is substantially larger due
to the increased number of valence electrons. This is particularly the case for the Bragg additivity rule based result (IAM-AR), and less so for IAM-PCM.
For $\rm Li^{3+}$ projectiles at the lowest energies shown this cross section for uracil is larger compared to that for methane by a factor of 1.5. This is in contrast
with the IAM-AR results, where this factor is about four. 
We are not aware of experimental data for these cross sections with $Q>1$, and therefore we cannot verify this aspect of the IAM-PCM prediction.

\begin{figure}
\begin{center}
\resizebox{0.6\textwidth}{!}{\includegraphics{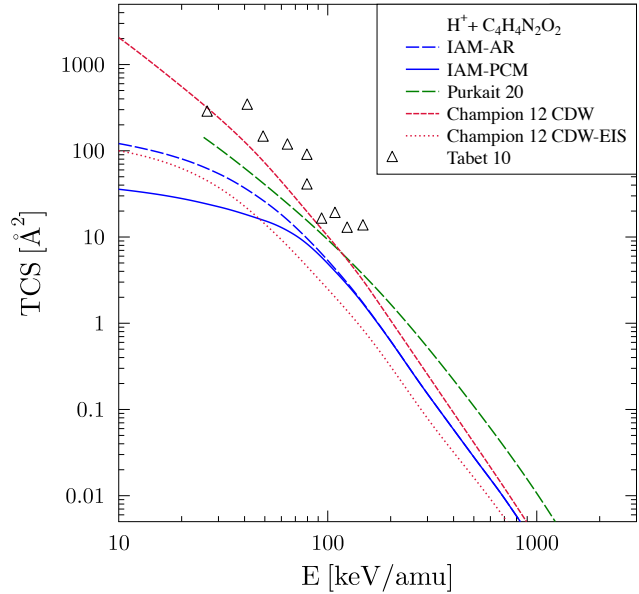}}  
   
\caption{Net electron capture cross section for collisions of protons with uracil ($\rm C_4 H_4 N_2 O_2$). 
The blue dashed curves show the IAM-AR results, while the blue solid curves are obtained with the IAM-PCM. 
The experimental data are from Ref.~\cite{PhysRevA.81.012711}, the other theoretical data are from Refs.~\cite{Champion_2012, Purkait_2019, Purkait_2020}
}
 \label{fig:Abb9}
\end{center}
\end{figure}

In Fig.~\ref{fig:Abb9} we compare our present results with the experimental data  and with other theoretical data for proton impact. The experimental data
of Tabet {\it et al.}~\cite{PhysRevA.81.012711} are higher than all theories shown for the energy range $40-150 \ \rm keV$. At energies above $100 \ \rm keV$ where both our IAM
calculations merge they fall in-between the CDW and CDW-EIS results of Champion {\it et al.}~\cite{Champion_2012}.
The distorted-wave CNDO calculations of Purkait{\it et al.}~\cite{Purkait_2019, Purkait_2020} cross the CDW results, while being lower below $100 \ \rm keV$
and higher above the transition point in energy where the IAM-AR and IAM-PCM results merge. At low energies the IAM-PCM results are clearly the lowest by about
a factor of three. 

\section{Conclusions}
\label{sec:conclusions}

As a follow-up to our previous work on net ionization of biologically relevant molecules by highly charged ions~\cite{PhysRevA.101.062709} we have presented
results for net capture for three molecules. The choice of water and methane was motivated by the existence of some experimental data, while uracil is an example
of a substantially larger molecule. Comparison with other theories shows that a number of them yield cross sections that are in the vicinity of Bragg additivity rule
results (IAM-AR).  For all molecules considered the IAM-PCM yields significantly reduced net capture cross sections at low energies with a 
 particularly strong effect for higher projectile charges.

For the water molecule two classical trajectory based simulations were used, the standard CTMC model was generally found to be closer to the IAM-AR results,
while the recently introduced time-dependent mean-field CTMC model at least partially supports the decrease in net capture cross sections provided by the IAM-PCM.

The IAM-PCM predicts that the net capture cross sections will saturate for high projectile charge by the presence of large contributions from constituent atoms
with large valence electron number. For small molecules the effect was demonstrated ($\rm C H_4$ cross sections are dominated by net cross sections from $\rm C$,
and likewise for $\rm H_2 O$ vs $\rm O$ targets),
while for uracil the cross sections are dominated by the leading atoms ($\rm C, N, O$). Future work may involve an investigation as to why many of the 
larger biomolecules have such similar cross sections, based on the notion that the dominant atoms ($\rm C, N, O, P$) located at their positions will control
the net cross sections through their effective geometric scattering cross sectional areas.

\begin{acknowledgments}
We would like to thank the Center for Scientific Computing, University of Frankfurt for making their
High Performance Computing facilities available.
Financial support from the Natural Sciences and Engineering Research Council of Canada (NSERC) is gratefully acknowledged. 
\end{acknowledgments}

%

\bibliography{ChargedProjectilesCapture}

\end{document}